%% file: manuscript.tex
\documentclass[twocolumn, tighten]{aastex631}
\input{commands}

\usepackage{booktabs}
\graphicspath{{./}{figures/}}

\shorttitle{LPVs in the LMC}
\shortauthors{Hey et. al}

\newcommand{\rev}[1]{\textcolor{black}{#1}}

\begin{document}

\title{The period-luminosity relation of long-period variables in the Large Magellanic Cloud observed with ATLAS}
\input{authors}

\begin{abstract}
    Period-luminosity relations of long period variables (LPVs) are a powerful tool to map the distances of stars in our galaxy, and are typically calibrated using stars in the Large Magellanic Cloud (LMC). Recent results demonstrated that these relations show a strong dependence on the amplitude of the variability, which can be used to greatly improve distance estimates. However, one of the only highly sampled catalogs of such variables in the LMC is based on OGLE photometry, which does not provide all-sky coverage. Here, we provide the first measurement of the period-luminosity relation of long-period variables in the LMC using photometry from the Asteroid Terrestrial-impact Last Alert System (ATLAS). We derive conversions between \textit{ugriz}, \textit{Gaia}, and ATLAS $c$ and $o$ passbands with a precision of $\sim 0.02$ mag, which enable the measurement of reliable amplitudes with ATLAS for crowded fields. We successfully reproduce the known PL sequences A through E, and show evidence for sequence F using the ratios of amplitudes observed in both ATLAS pass-bands. 
    Our work demonstrates that the ATLAS survey can recover variability in evolved red giants and lays the foundation for an all-sky distance map of the Milky Way using long-period variables.
\end{abstract}


\section{Introduction}

\begin{figure*}[t]
    \includegraphics[width=\linewidth]{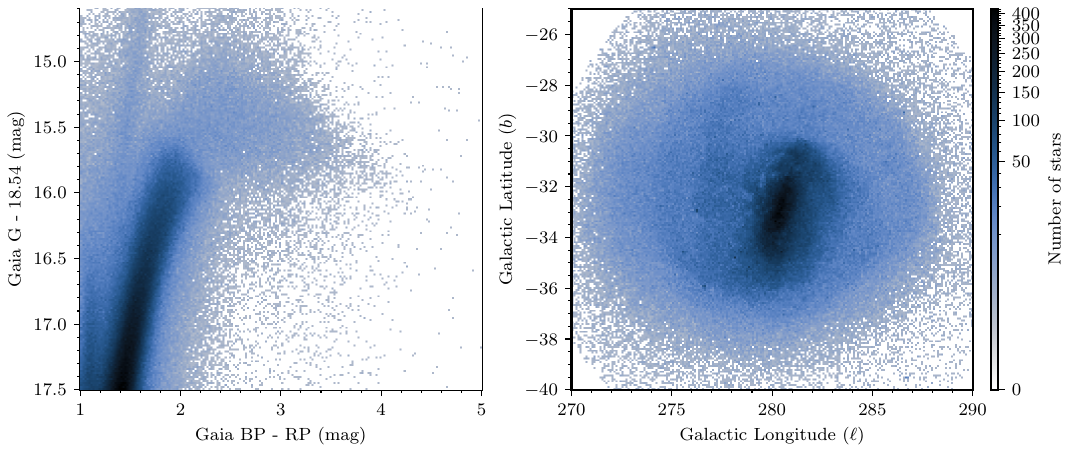}
    \caption{\textit{Gaia} DR3 density plots of evolved stars around and within the LMC following the cuts made in Sec.~\ref{sec:cuts}. Left: the color-magnitude diagram of targets after removing contaminant foreground stars, showing the red giant branch. Right: Same as left panel, but showing the galactic coordinates for each source.}
    \label{fig:magellanic}
\end{figure*}

Long-period variables (LPVs) are evolved, cool red-giants typically sitting in the M-type spectral class. Their variability is ubiquitous, because essentially all stars on the red giant and asymptotic giant branch are intrinsically variable. 
LPVs are generally divided into semiregular variables (SRVs) and Mira variables, based on the regularity and amplitude of their light curves \citep{Lopes2015Variability}. The periods observed in LPVs follow a series of distinct, parallel period-luminosity (PL) sequences that correspond to different pulsation modes, with the most well-known sequence belonging to the Mira variables. Their study has been greatly accelerated by ground-based large-scale variability surveys, such as MACHO \citep{Wood1999MACHO}, OGLE \citep{Soszynski2004Optical, Soszynski2009Optical}, ASAS-SN \citep{Jayasinghe2018ASASSN, Auge2020Gaia}, as well as space-based missions like Hipparcos \citep{Bedding1998Hipparcos, Tabur2009Longterm, Tabur2010Periodluminosity}, CoRoT \citep{Lopes2015Variability}, Kepler \citep{Banyai2013Variability}, and most recently, Gaia \cite{Lebzelter2022Gaia, Lebzelter2023Gaia}.

Large-scale variability surveys greatly improved our knowledge of the properties of LPVs. \citet{Minniti1998Pulsating} presented preliminary results of the search for semiregular variables (SRVs) in the bulge using the photometric data collected by the MACHO project. They reported the discovery of about 2000 SRVs which followed two parallel sequences in the period–color diagram. \citet{Alard2001Masslosing} used MACHO photometry to study the mass loss in about 300 AGB stars in the Galactic bulge. \citet{Glass2003Periodmagnitude} extended this sample to more than 1000 objects and showed a series of four sequences in the log-period and $K$-magnitude plane – similar to those discovered by \citet{Wood1999MACHO} in the Large Magellanic Cloud (LMC).

A realization from Kepler photometry was that the PL relations in SRVs are a simple extension of the pattern of radial and non-radial solar-like oscillations observed in lower-luminosity red giants \citep{Stello2014Nonradial}, which follow a scaling between oscillations periods and amplitudes \citep{Huber2011Solarlike, Banyai2013Variability, Mosser2013Periodluminosity, Yu2020Asteroseismology}. In the classical approach, SRV stars are distinguished from Mira variables by their visual peak-to-peak light amplitudes of less than 2.5 magnitudes. However, there is no natural break in the amplitudes between these two types of long-period variables, especially since many of them tend to change their amplitudes, so the same star may be classified as a Mira or as an SRV, depending on the observational time span. The Mira sequence in the PL relation has been used to probe the structure of the Milky Way \citep{Iwanek2022Threedimensional}. The potential of other sequences comprising the SRVs as viable distance indicators has been investigated to some degree, but has only recently begun to gain traction \citep{Trabucchi2017New, Trabucchi2019Characterisation, Trabucchi2021Semiregular, Hey2023Far}.

One of the key difficulties in using the PL relation for measuring distance relates to the fact that they are adjacent and nearly overlapping in period-magnitude space. Thus, a single period of oscillation is insufficient to determine the true sequence, leading to an ambiguous absolute magnitude and thus large uncertainty in the inferred distance. \citet{Tabur2010Periodluminosity} demonstrated that this difficulty can be overcome by using multiple periods of oscillation, assuming that \rev{the majority of sequences on the PL relation are associated to different radial orders of the same angular degree, as discussed by \citet{Mosser2013Periodluminosity, Trabucchi2019Characterisation, Trabucchi2021Semiregular}.} By calculating the probability of each period belonging to each sequence, it is possible to uniquely determine the magnitude and calculate a distance modulus precise to at least 15\%. When combined with the amplitude of the pulsation, the sequences are further well-separated. This technique was refined and applied in \citet{Hey2023Far} to determine the distances to approximately 200,000 stars in and around the Galactic bulge, allowing the authors to probe the kinematics of the Galactic center to an unprecedented degree. This was accomplished using OGLE photometry of Galactic bulge stars, with distances calculated using the OGLE LMC relation. More recently, \citet{Zhang2024Kinematics} used the \textit{Gaia} low-amplitude long-period variables (the SRVs) to measure distances and kinematics around the Galactic center. Their method involved using amplitudes to distinctly separate out the PL sequence associated with these variables, compared to the holistic approach in \citet{Hey2023Far} where each sequence was treated in the same manner.

Regardless of the method, it is now evident that an all-sky analysis at the same precision requires the amplitudes of the PL relation to be measured (or converted) into the appropriate pass-band. The highest quality PL relation of the LMC is based on data from the OGLE survey with photometry in the $I$- and $V$-bands. The amplitudes of the LPVs in the $I$- and $V$-band can not be easily translated into other photometric systems, because the amplitudes and periods are prone to change depending on the observational time-span (i.e., a single magnitude measurement on a long-period Mira variable can differ by over a magnitude depending on the pulsation phase). Likewise, there is a well-known wavelength dependence of pulsation amplitudes, which decrease towards longer wavelengths \cite{Aerts2010Asteroseismology}.

In this paper, we measure the LMC PL relation using light curves from the Asteroid Terrestrial-impact Last Alert System (ATLAS; \citealt{Tonry2018ATLASa}), a ground-based all-sky survey with near daily cadence. We select LMC candidate members with \textit{Gaia} DR3 astrometry, and calculate forced differential photometry light curves for each candidate \rev{(Sec.~\ref{sec:atlas})}. Using these light curves, we provide relations to translate from the OGLE $I$- and $V$-band passbands, \textit{Gaia}, and $ugriz$ photometry into the ATLAS orange ($o = r + i$) and cyan ($c = g+r$) filters to account for crowded ATLAS fields \rev{(Sec.~\ref{sec:calibration})}. We use an automated prewhitening routine to calculate all frequencies and amplitudes of periodicity in each light curves to a 1\% significance level \rev{(Sec.~\ref{sec:prewhitening})}. From this, we re-derive the LMC PL relation and recover the typical known sequences A through E. We make available all ATLAS light curves used in this paper to facilitate future analysis.

\section{Data}

\subsection{LMC membership}
\label{sec:cuts}

The initial step in our selection process involves extracting a spatially defined sample from the \textit{Gaia} Data Release 3 (DR3) catalog. We centered this sample around the established position of the LMC ($\alpha$, $\delta$) = (81.28$^\circ$, -69.78$^\circ$) \citep{Marel2002New}), encompassing a radius of 10$^\circ$. We further applied additional constraints based on \textit{Gaia} photometric data. Specifically, we implemented cuts on the \textit{Gaia} color index ($G_{\rm bp}$ - $G_{\rm rp}$) to isolate cool stars, which are more likely to be giants. Furthermore, we utilized the well-established distance modulus to the LMC (18.476 mag; \citealt{Pietrzynski2019Distancea}) to apply an absolute magnitude cut, effectively selecting for giant stars at the approximate distance of the LMC. The ADQL query used to perform this selection is reproduced below:

\noindent {\tt \small
    SELECT g.source\_id, g.parallax, ...\newline
    FROM gaiadr3.gaia\_source as g\newline
    WHERE 1=CONTAINS(POINT('ICRS',g.ra,g.dec), \newline
    CIRCLE('ICRS',81.28,-69.78,10))\newline
    AND (g.phot\_g\_mean\_mag - 18.54) < -1\newline
    AND (g.phot\_g\_mean\_mag - 18.54) > -4\newline
    AND g.bp\_rp > 1.\newline
    AND g.bp\_rp < 5.5\newline
}

The initial selection process yielded a sample of 932,513 objects. To refine this dataset and minimize contamination from foreground Milky Way stars, we implemented a simplified version of the approach described by \citet{Luri2021Gaia} and \citet{GaiaCollaboration2018Gaia}. Our methodology focuses on proper motion selection and astrometric quality cuts. We provide a brief overview of our procedure below, and refer readers seeking a more comprehensive approach to \citet{Luri2021Gaia}.

Our refinement process consisted of the following steps:

\begin{enumerate}
    \item We excluded stars with fractional parallax uncertainties $\varpi/\sigma_{\varpi} < 5$. This criterion effectively eliminates stars with high-precision parallax measurements, which are more likely to be nearby Milky Way objects rather than LMC members.

    \item We applied a limiting magnitude cut of $G < 19.5$. This threshold serves two purposes: it excludes sources with less precise astrometry, which are typically fainter, and ensures compatibility with the ATLAS photometry described in Section~\ref{sec:atlas} by removing light curves with high flux uncertainties.

    \item At this stage, our methodology diverges from that of \citet{Luri2021Gaia}. We calculated the total proper motion ($\mu = \sqrt{\mu_\alpha^2 + \mu_\beta^2}$) for each remaining star. Subsequently, we retained only those stars with proper motions within two standard deviations of the mean, corresponding to a 95\% confidence interval.
\end{enumerate}

\noindent Our decision to apply a more simplified approach to membership selection is justified by two primary considerations. Firstly, we prioritize the accumulation of as many LMC giant star candidates as possible, even at the risk of an increased false-positive rate. Secondly, any residual contaminant stars in our selection can be more readily identified and removed during the construction of the period-luminosity relation, as their measured variability periods and amplitudes will not align with the known sequences.

The selection criteria resulted in a sample of 631,259 objects for the LMC. The spatial distribution of these objects is illustrated in Figure~\ref{fig:magellanic}.  We cross-match this sample with the 2MASS Point Source Catalog (2MASS; \citealt{Cutri20032MASS, Skrutskie2006Two}) with a matching 5 arcsecond radius to obtain $J$, $H$, and $K$-band magnitudes of the sources.

\subsection{OGLE photometry and supplementary data}

The Optical Gravitational Lensing Experiment  (OGLE; \citealt{Soszynski2011Optical, Soszynski2009Optical, Soszynski2013Optical}) is a ground-based high-resolution variability survey across the Milky Way and its immediate environs. The primary goal of the mission is the characterization and detection of lensing sources across the sky (e.g., \citealp{Suzuki2018Likely, Han2018OGLE2017BLG0039, Poleski2018Ice, Mroz2024Microlensing}), which has also enabled various adjacent science results, including eclipsing binaries \cite{Graczyk2018Latetype,Suchomska2019Accurate} and variable stars \cite{Udalski2018OGLE, Pilecki2018Dynamical}.

The third OGLE data release contained photometry and classifications for semiregular and long-period variables in the Large and Small Magellanic Clouds \citep{Soszynski2009Optical, Soszynski2011Optical}. They presented 91,995 long-period variable candidates in the LMC, of which 79,200 were classified as the so-called OGLE Small Amplitude Red Giants (OSARGS; \citealt{Wray2004OGLE}) \footnote{Briefly, OSARGS are defined as RGB or AGB stars with pulsation periods from 10 to 100 days and amplitudes of a few millimagnitudes. This definition is defined solely to separate them from higher amplitude variables, hereafter referred to as semiregular variables.}, 11,128 as semiregular variables, and 1,667 Mira variables. The catalog includes the dominant, secondary, and tertiary periods of variability determined by iterative fitting and subtraction of a third-order Fourier series, as well as their peak-to-peak amplitudes, and multi-epoch $I$- and $V$-band photometry. The dominant periods of variability were manually classified, providing an extremely clean sample of LPVs in the Magellanic clouds. We use the catalog, which we cross-match to Gaia DR3, as well as the $I$-band light curves, which were used for reporting amplitudes.

\subsection{ATLAS photometry}
\label{sec:atlas}

\begin{figure}[t]
    \centering
    \includegraphics[width=\linewidth]{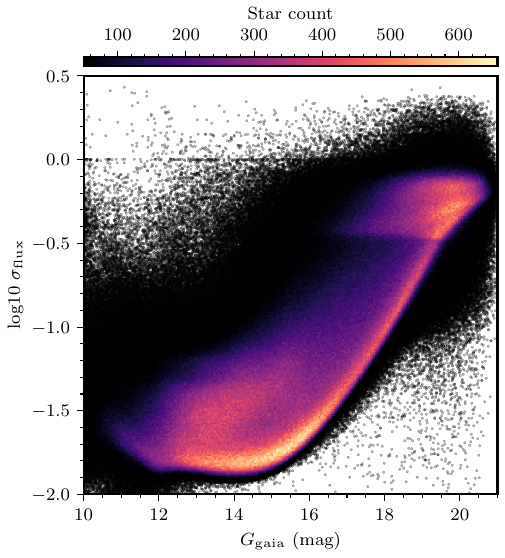}
    \caption{Sensitivity of the ATLAS photometry to a series of 5 million randomly sampled M-giant stars across the sky (both Northern and Southern hemispheres). The y-axis indicates the scatter of the flux across the entire light curve, and the x-axis is the apparent Gaia DR3 G-band magnitude.}
    \label{fig:sensitivity}
\end{figure}

ATLAS \citep{Tonry2018ATLAS} is an all-sky survey designed to find asteroids on their final approach towards Earth. Its primary purpose as a last alert mission requires near nightly-cadence observations of the entire sky. ATLAS incorporates a robotic 0.5m f/2 Wright Schmidt telescope with a 5.4 x 5.4 degree field of view and 1.86" pixel scale. The ATLAS project has been comprehensively monitoring the night sky since 2017. Initially covering the area from the north celestial pole to -50° declination every two nights, the system expanded in 2022 with two Southern Hemisphere telescopes, now enabling nightly observation of the entire dark sky. The all-sky high-cadence observations performed by ATLAS make it a prime resource for the discovery and characterization of bright long-period variable stars, such as Cepheid and Mira variables. Indeed, one such ATLAS variability study identified hundreds of thousands of candidate variable stars \citep{Heinze2018First}.

We use the ATLAS forced photometry light curves -- differential flux light curves calculated at pre-defined positions.
In the ATLAS forced photometry, a point-spread-function is calculated for each image based on high signal-to-noise stars, and a profile fit is forced at the input coordinates. The forced measurement is made on a difference image, which helps suppress artifacts from crowding but compels us to ascertain the mean brightness of the giant by other means.  Data processing and photometry are described in more detail in \citet{Tonry2018ATLAS} and \cite{Smith2020Designa}.

Included in the forced photometry are data quality flags, flux uncertainties, and a per flux measurement of the reduced $\chi^2$ of the PSF fit. For all ATLAS light curves used throughout this paper, we ensure that the fractional flux uncertainty is less than 20\%, the quality flag is `good', and the reduced $\chi^2$ is less than 10. We also apply a 3$\sigma$ outlier rejection, as well as a weak Gaussian filter to remove extreme discontinuities in the flux measurements.

As a test inspection of the ATLAS photometry, we extracted light curves for 5 million cool (Gaia $G_{bp}$ - $G_{rp}$ $\geq 1.3$ mag) random giant stars. We use giant stars to probe photometric properties consistent with our LMC sample. In the absence of low uncertainty parallaxes for the giants, we separate them from their dwarf star counterparts with a reduced proper motion cut in Gaia (RPM; \citealt{Koppelman2021Reduced}). Likewise, we also select stars with an apparent magnitudes less than 21.5 in Gaia $G$. From this sample, we measure the standard deviation of the photometric flux for both bands combined in Fig.~\ref{fig:sensitivity}. This demonstrates some important characteristics of the ATLAS data. In particular, the flux precision rapidly decreases for sources fainter than 20th magnitude in the Gaia $G$-band, while the ideal spot for observations seems to be around 15th magnitude. The precision floor is caused by the difficulties in flattening images over 30~deg$^2$ (20$\times$ OGLE-IV) in all weather conditions, and no requirement to do better in order to find asteroids. Saturation becomes an issue at around 10th magnitude and brighter. Note that this sample includes sources from both the Northern and Southern hemispheres, which have different observing baselines.

\section{Calibration of ugriz and Gaia passbands to ATLAS}
\label{sec:calibration}

\begin{table}
    \centering
    \caption{Photometric pass-band conversions for the ATLAS data.}
    \begin{tabular}{ll|lll|ll}
        \hline
        Y              & X                   & a$_1$  & a$_2$  & a$_3$  & $\sigma_{\rm 68th}$ & $\sigma_{\rm 95.4th}$ \\
        (mag)          & (mag)               &        &        &        & (mag)               & (mag)                 \\
        \hline
        ($o$-$r$)      & ($r$-$i$)           & -0.307 & -0.071 & -0.087 & 0.012               & 0.048                 \\
        ($o$-$i$)      & ($r$-$i$)           & 0.585  & 0.015  & -0.055 & 0.009               & 0.042                 \\
        ($c$-$r$)      & ($g$-$r$)           & 0.189  & 0.098  & 0.121  & 0.01                & 0.053                 \\
        ($c$-$g$)      & ($g$-$r$)           & -0.958 & 0.162  & 0.200  & 0.014               & 0.054                 \\
        ($o$-$G_{bp}$) & ($G_{bp}$-$G_{rp}$) & -0.433 & -0.057 & 0.130  & 0.005               & 0.068                 \\
        ($o$-$G_{rp}$) & ($G_{bp}$-$G_{rp}$) & -0.251 & 0.178  & 0.559  & 0.006               & 0.023                 \\
        ($c$-$G_{bp}$) & ($G_{bp}$-$G_{rp}$) & -0.563 & 0.156  & 0.240  & 0.007               & 0.030                 \\
        ($c$-$G_{rp}$) & ($G_{bp}$-$G_{rp}$) & 0.714  & 0.069  & 0.029  & 0.004               & 0.030                 \\
        \hline
    \end{tabular}
    \label{tab:conversion}
\end{table}

\begin{figure*}
    \includegraphics[]{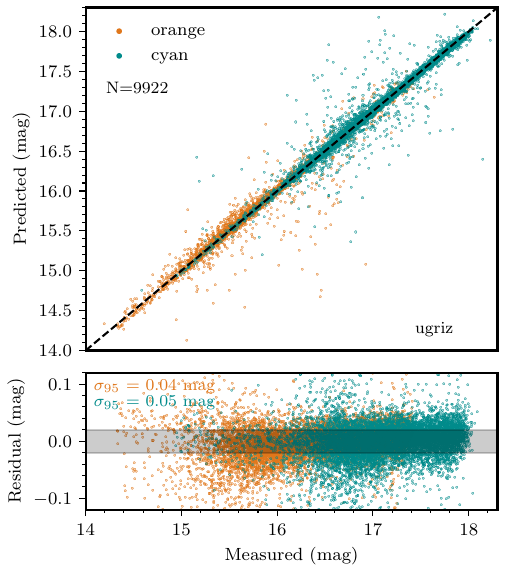}
    \includegraphics[]{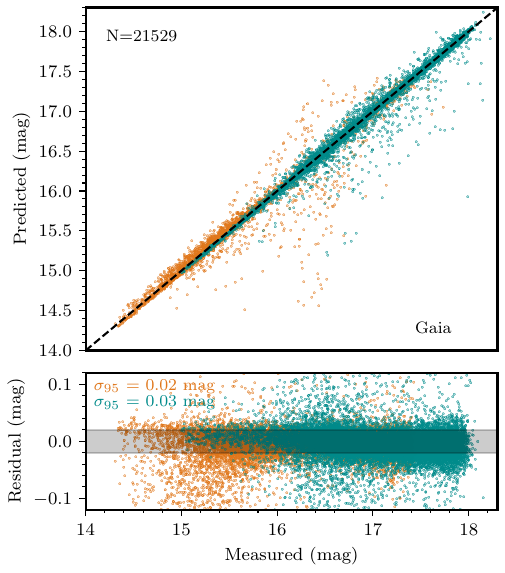}
    \caption{Comparison of measured and predicted orange and cyan ($o$/$c$) band magnitudes using the relations provided in Eq.~\ref{eq:conversion} with constants defined in Table~\ref{tab:conversion} for the $ugriz$ (left) and Gaia (right) passbands. The bottom panel shows the residual magnitudes, with the gray band highlighting the 0.02 mag nominal ATLAS photometric precision.}
    \label{fig:conversions}
\end{figure*}

\begin{figure*}[t]
    \centering
    \includegraphics{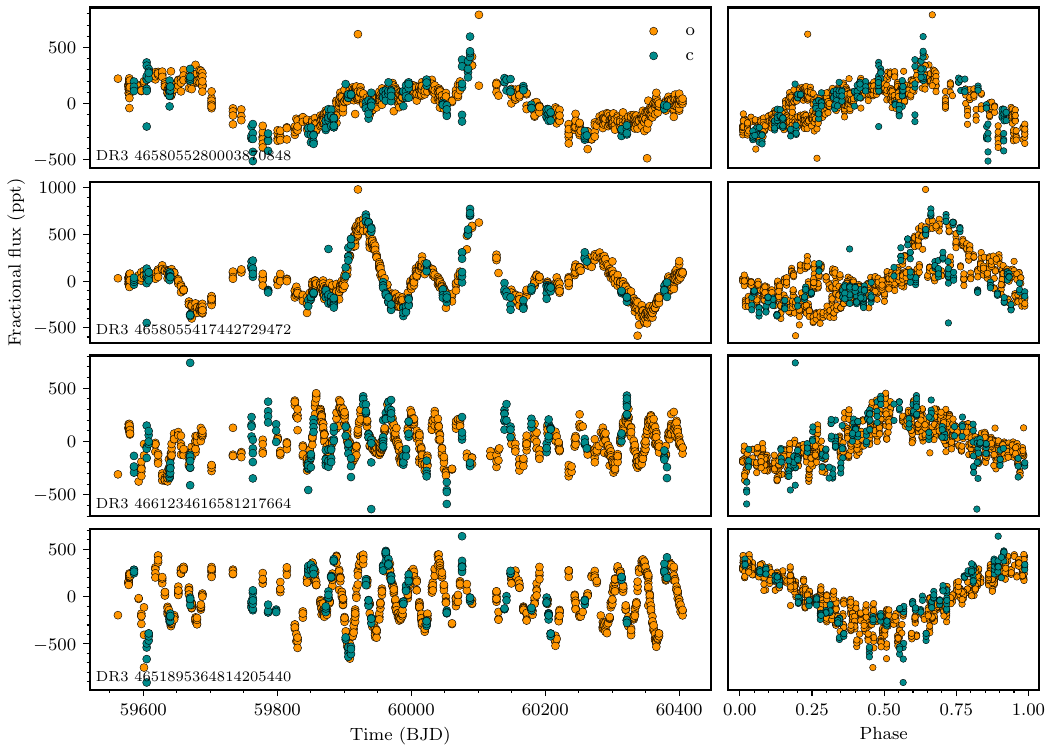}
    \caption{Examples of four randomly chosen variables in our sample, where the flux is corrected following Sec.~\ref{sec:atlas}. The left panels show the full ATLAS light curve in both pass-bands (orange, cyan), while the right panels show the light curve folded on the dominant measured period.}
    \label{fig:example_lc}
\end{figure*}

The ATLAS survey, with a pixel scale of 1.86", presents a unique challenge when observing dense stellar environments such as the LMC. In these crowded fields, individual stars are likely to be contaminated by flux from nearby sources, resulting in an artificially elevated base flux. This contamination is particularly pronounced in regions of high stellar density. To accurately measure the amplitude of stellar variability, it becomes necessary to convert from a higher-resolution photometric system to the ATLAS magnitude system, which utilizes two primary bands: orange ($o$; 560-820 nm) and cyan ($c$; 420-650 nm). It is important to note that we cannot rely on the conversion relations provided by \citet{Tonry2018ATLAS} in the Stellar Reference Catalog (REFCAT) for transformations between Gaia and 2MASS passbands to the $ugriz$ system. These relations were derived using main sequence stars and less cool giant stars than the population discussed here.

Although the REFCAT catalog has uniform $ugriz$ magnitudes for each star discussed in this sample, the sources for its photometry vary across the sky. $ugriz$ photometry in REFCAT for the Southern Hemisphere were taken from the SkyMapper DR2 catalog \cite{Onken2019SkyMapper}, whose values are biased due to crowding in the LMC. To avoid this problem, we cross-matched our sample against the Survey of the Magellanic Cloud (SMASH; \citealt{Nidever2021Second}) using a 5 arcsecond matching radius. SMASH is a high-resolution survey of the Magellanic Clouds observed using the Dark Energy Camera (DECam) in the $ugriz$ passbands. We find that approximately 80\% of our sample could be reliably cross-matched to SMASH.

To calibrate into the ATLAS orange and cyan bands, we made several quality cuts on our photometric data. Most importantly, we ensured the stars were not contaminated by nearby stars. The contamination was selected through the use of the {\tt rp1} parameter in REFCAT. {\tt rp1} is a proximity statistic derived by summing the cumulative $G$-band flux of all \textit{Gaia} stars as a function of distance from each star and reporting the radius where this flux first exceeds 0.1. We ensure {\tt rp1} is greater than five arcseconds. We also only use photometry with at least 200 flux measurements, and ensure the fractional uncertainty in colors are less than 10\%. From these cuts, for each star, using the reduced photometry, we calculate the mean magnitude in both bands after typical sigma clipping and light curve cleaning. We then fit an equation of the form
\begin{equation}
    Y= a_1 X + a_2 X^2 + a3
    \label{eq:conversion}
\end{equation}
where $X$ and $Y$ are photometric colors (i.e., ($G_{bp}$ - $G_{rp}$)), and $a{1,2,3}$ are the coefficients of the polynomial. 

We provide pass-band conversions between Gaia \textit{BP/RP}, and \textit{ugriz} to the ATLAS orange and cyan bands in Table~\ref{tab:conversion} and show the conversions for a subset of these relations for both Gaia and \textit{ugriz} photometry in Fig.~\ref{fig:conversions}. For the remainder of this paper, we use the conversions with the lowest residual scatter. That is, ($o$-$G_{rp}$) and ($c$-$G_{rp}$). We caution that these relations are only valid for cooler stars ($G_{bp}$ - $G_{rp}$ > 1.4), and users wishing to calibrate photometry for hotter stars should refer to \citet{Tonry2018ATLAS}.

Using these relations allow us to calculate the expected flux of the star measured by ATLAS in Janskys;
\begin{equation}
    \mu Jy = 10^{-0.4 (m_{\rm AB} - 23.9)},
\end{equation}
where $m$ is the predicted magnitude in the corresponding orange or cyan band. In practice, for each ATLAS light curve, \rev{we calculate the predicted flux in both bands and use that as the baseline flux. We then divide the flux by the median value to obtain the light curve in fractional flux units, which we use for the remainder of the paper. These fractional units are multiplied by 1,000 and are given as parts-per-thousand (ppt).} We show an example of several ATLAS light curves in the LMC sample in Fig.~\ref{fig:example_lc}.

\section{The Period-luminosity relation}
\subsection{Prewhitening}
\label{sec:prewhitening}

\begin{figure*}[t]
    \centering
    \includegraphics{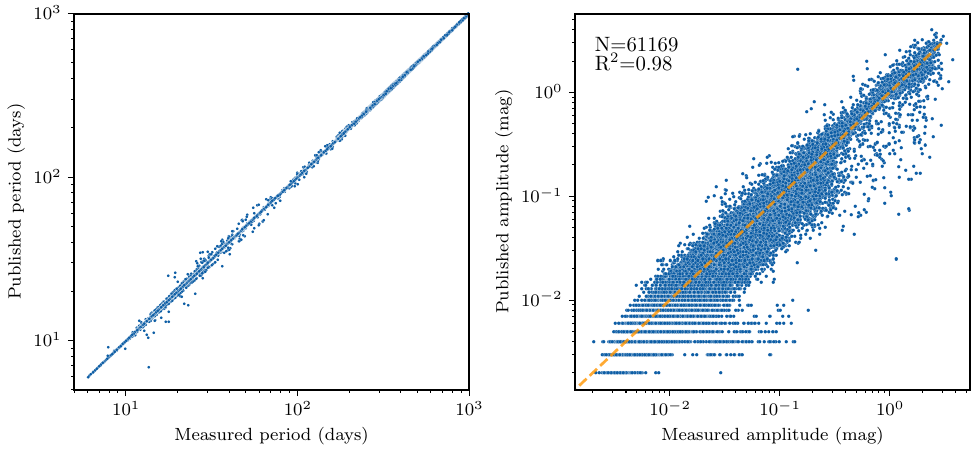}
    \caption{Comparison of dominant periods (left) and amplitudes (right) measured from the OGLE $I$-band photometry (x-axis) and compared against the published catalog (y-axis). Note that the published amplitudes appear to be measured as peak-to-peak amplitudes, and are thus divided by two to match the semi-amplitudes measured here. The discretization of amplitudes in the published catalog is likely due to the non-fitting nature of their prewhitening routine.}
    \label{fig:ogle_compare}
\end{figure*}

To analyze candidate pulsators in our sample, we employ an automated prewhitening scheme based on the approach used for main-sequence variable stars in \citet{Hey2021Searcha} and further developed in \citet{Hey2024Confronting}. This method iteratively selects the highest peak in the amplitude spectrum above a 1\% false alarm level \citep{VanderPlas2018Understanding}, fits a sinusoid characterized by frequency, amplitude, and phase, and subtracts it from the light curve. This process continues until no peaks remain above the 1\% false alarm probability level.

Our approach differs from previous studies of semiregular variables. The OGLE mission surveys use a third-order Fourier series fit \citep{Soszynski2004Optical}, which is more susceptible to over-fitting and spurious peak identification. \citet{Tabur2009Longterm} use a similar method to ours but with a stopping criterion based on the signal-to-noise ratio of peaks, as in \citet{Frandsen1995CCD}. After testing various stopping conditions for identifying variability in long-period variables, we found the false alarm probability to be the most reliable for recovering periodicity. While this may result in more false positives, these are removed in post-processing (see Sec.~\ref{sec:kde}). For a comprehensive discussion of prewhitening strategies, we refer readers to \citet{Beeck2021Detection}.

\begin{figure}[t]
    \centering
    \includegraphics[width=\linewidth]{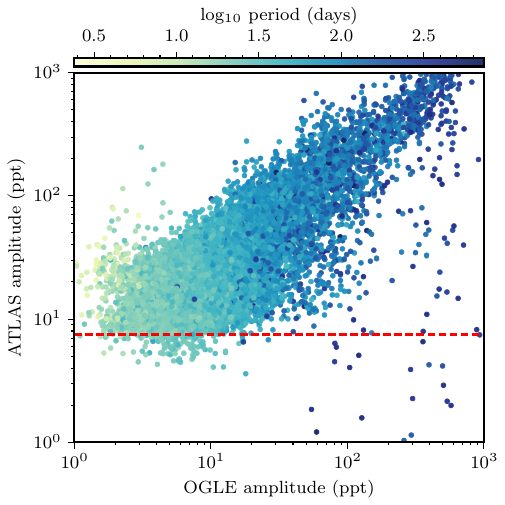}
    \caption{Comparison of amplitudes for the same period in the OGLE and ATLAS data. The red dashed line marks the average lower bound of the ATLAS amplitude measurements, around 7.6 ppt.}
    \label{fig:amp_compare}
\end{figure}

\begin{figure*}
    \centering
    \includegraphics{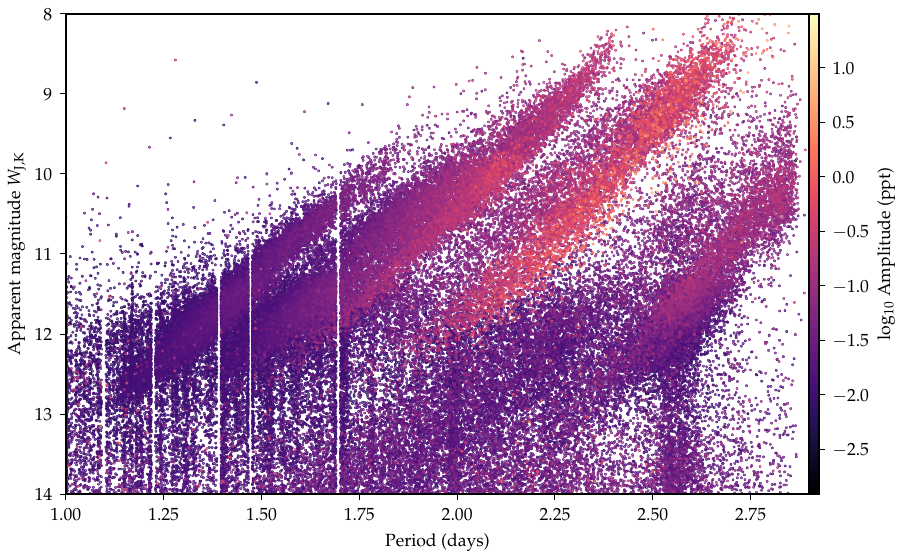}
    \includegraphics{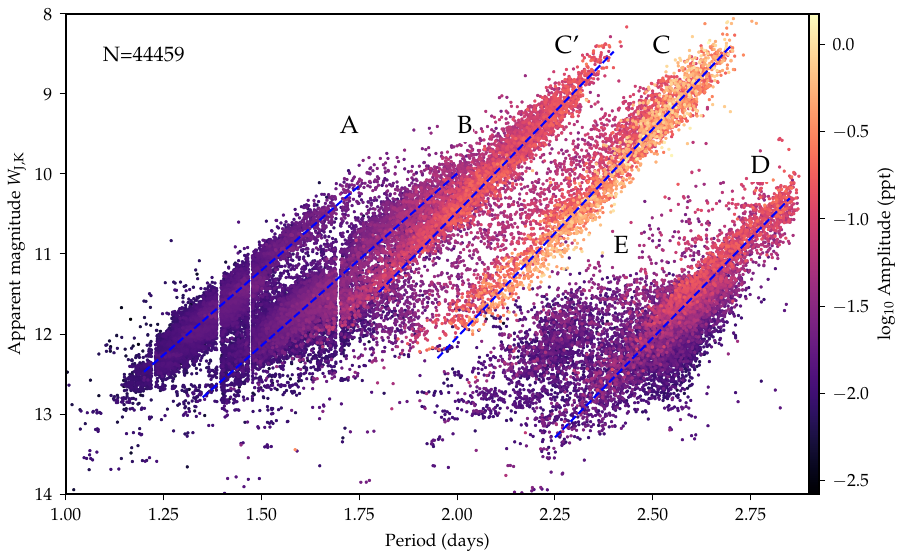}
    \caption{Top: The period-luminosity relation for the stars in the LMC sample with amplitude as measured in the ATLAS $o$-band. We show only the dominant period of variability, which we define as that of the highest amplitude. Note the white strips comprising removed systematics near the lunar synodic period ($\sim$29 days) and integer multiples thereof. Bottom: The same PL relation after only selecting high-probability members as discussed in Sec.~\ref{sec:kde}. Note that this selection still retains a handful of systematic contaminants at short periods. Individual sequences are fit as discussed in the text.}
    \label{fig:pl}
\end{figure*}

To validate our prewhitening method, we first applied it to the original OGLE $I$-band photometry, comparing our results against the published catalog values from \citet{Soszynski2013Optical}.  Fig.~\ref{fig:ogle_compare} shows this comparison for LMC stars. We found 67\% agreement between measured and published dominant periods, within a 10-day tolerance. We consider this an acceptable amount given the manual adjustments in the OGLE catalog.

Interestingly, we observed significant differences in measured mode amplitudes despite good period agreement. This discrepancy likely stems from differing amplitude definitions: the catalog uses a third-order Fourier series fit, while we employ a first-order fit. Additionally, OGLE reports peak-to-peak amplitudes, whereas we use semi-amplitudes as defined in \citet{Kjeldsen1995Amplitudes}. For instance, a sinusoid with amplitude 1 has a peak-to-peak amplitude of 2 and a semi-amplitude of 1 in the amplitude spectrum.

We then compared OGLE and ATLAS light curves for the same targets, noting that our Gaia-based LMC selection encompasses the entire OGLE selection. To facilitate amplitude comparison, we converted the OGLE $I-$band magnitudes to flux density using the standard relation and Johnson zero-point; $F = (2380)10^{-m / 2.5}$.

ATLAS light curve fluxes were corrected as described in Sec.~\ref{sec:calibration}. We prewhitened both datasets using identical methods, measuring the first 100 points with a false alarm level below 1\%. Most stars exhibited fewer than 3 peaks. Fig.~\ref{fig:amp_compare} illustrates this comparison for stars with period differences less than 1 day, using OGLE $I$-band and ATLAS o-band data. Generally, ATLAS amplitudes tend to be higher than OGLE, likely due to the bluer ATLAS passband where amplitudes of variable stars are typically measured to be larger.

Finally, we applied the prewhitening routine to all light curves in our sample for periods between 2 and 790 days (the maximum ATLAS timespan). We treated ATLAS orange and cyan pass-bands as separate light curves. Of the 631,529 LMC targets, 379,920 yielded at least one frequency measurement above the 1\% false alarm probability threshold in either the orange or cyan passbands. The remaining targets had no discernible variability.

\subsection{Period selection}
\label{sec:kde}
% ORIGINAL
\rev{Figure \ref{fig:pl} shows the period-luminosity (PL) diagram of the LMC sample in the ($W_{\rm J K_s}$, $\log{P}$) plane, where $W_{\rm J K_s}$ is the Wesenheit index defined as \begin{equation}
    W_{\rm J K_s} = K_s - 0.686 (J - K_s).
\end{equation}}
We observe a series of spurious peaks in the 379,920 targets measured in the ATLAS photometry. \rev{Most obvious of these is the systematic of the lunar synodic period, as well as frequencies associated with the daily cadence of observations which are removed with a direct cut in period.} In the absence of a longer time-span of observations, it becomes difficult to identify which frequencies truly belong to the sequences of the PL relation, and which are spurious. \rev{Furthermore, a dominant source of scatter in the PL relation is caused by the single-epoch observations using 2MASS photometry. Since the stars are variable, a single epoch magnitude measurement might not represent the mean apparent magnitude of the star.}

We use the original OGLE PL relation to remove outliers from the ATLAS PL relation. We first cross-match the ATLAS sample to the OGLE sample, and find matching periods within 1 day. We then calculate a Kernel Density Estimate (KDE; \citealt{Jones2001SciPy, Virtanen2019SciPy}) in period, amplitude, and magnitude space, similar to that performed in \citet{Hey2023Far}. \rev{This yields a non-parametrized estimate of the probability distribution of the PL relation. Using this, we can query the log-likelihood for any arbitrary point on the KDE. We calculate this for each set of measurements (period, amplitude, magnitude) in our prewhitened ATLAS data.} This method works especially well for removing systematics present in the original sample (Fig.~\ref{fig:pl}). However, it requires the ATLAS PL to match the OGLE PL, despite some differences in the distributions. We summarize the LMC prewhitening results in Table~\ref{tab:results}.

We retain prewhitened values with a log-likelihood greater than -20, an arbitrary choice which recovers the sequences without rejecting too many values. Indeed, visual inspection of the distribution of the log-likelihoods for the sample reveals a strongly bimodal structure splitting at -20, which we infer as real members of the PL relation and outliers at lower probabilities. The lower panel of Figure \ref{fig:pl} shows the PL relation for the corrected sample. The corrected sample consists of 51,238 unique targets, 44,459 of which are measured in the `o' band, and 15,558 of which are measured in the `c' band. 

In the PL relation shown, the parallel sequences A, B, C, and D correspond to different pulsation modes of different spherical degree and radial order  \citep[e.g.][]{Wood1999MACHO, Yu2021Asteroseismology}, while sequence E consists of suspected binary systems. The PL relations show a strong correlation with pulsation amplitude, with more luminous and longer period stars having higher amplitudes than their more rapidly variable counterparts, as predicted by \citet{Stello2009relation}.

\begin{table}[]
\begin{tabular}{lllll}
\toprule
& $W_{J, K} = \alpha (logP - 2.0) + \beta $ \\ 
Sequence   & $\alpha$           & $\beta$           &  &  \\
   \midrule
A  & -4.23($\pm0.02$)  & 9.10($\pm0.01$)  &  &  \\
B  & -4.30($\pm0.02$)  & 9.99($\pm0.02$) &  &  \\
C' & -5.01($\pm0.03$) & 10.475($\pm0.005$) &  &  \\
C  & -5.21($\pm0.02$) & 12.04($\pm0.01$) &  &  \\
D  & -4.97($\pm0.07$)  & 14.53($\pm0.05$) &  & \\
\bottomrule
\end{tabular}
\caption{\label{tab:pl_fit} Fits to individual sequences of the PL relation for the sample in the Wesenheit index.}
\end{table}

\rev{Compared to the PL relation measured by \citet{Soszynski2007Optical} using OGLE, the relation presented here is noisier due to the shorter observational time-span of the ATLAS data, and we are unable to observe sub-sequences previously identified. Despite this, we still measure the relations of individual sequences A through E following the standard equation, but refer the reader to \citet{Soszynski2007Optical} for more detailed sequence analysis. To measure the sequences, we make several cuts along the amplitude and period space, followed by a K-means clustering algorithm to separate out individual sequences. We then fit a linear equation of the form
\begin{equation}
    W_{\rm J,K} = \alpha (\log{P} - 2.0) + \beta,
\end{equation}
using non-linear least squares, and summarize the results of this fit in Table~\ref{tab:pl_fit}. Note that we make no distinction here for the Oxygen and Carbon rich Mira variables comprising sequence C` and C. In practice, their sequences are weakly differentiated by a steeper relation for the C-rich Mira variables. We also do not fit for sequence E, thought to comprise ellipsoidal variables.}

\begin{table*}
    \centering
    \begin{tabular}{llrrrrrrrrr}
        \toprule
        Gaia DR3 ID         & camera & K$_{\rm mag}$ & RA    & Decl.  & P$_{1}$ & P$_{2}$ & P$_{3}$ & A$_{1}$ & A$_{2}$ & P$_{3}$ \\
                            &        & (mag)         & (deg) & (deg)  & (d)     & (d)     & (d)     & (ppt)   & (ppt)   & (ppt)   \\
        \midrule
        5283967220539719936 & o      & 13.55         & 92.26 & -66.86 & 74.60   & --      & --      & 8.99    & --      & --      \\
        5283967220539719936 & c      & 13.55         & 92.26 & -66.86 & 3.40    & 2.15    & --      & 24.36   & 17.49   & --      \\
        4658045418756865792 & c      & 13.33         & 82.57 & -69.40 & 4.78    & --      & --      & 26.19   & --      & --      \\
        4658037550374108288 & o      & 12.74         & 82.98 & -69.43 & 15.39   & --      & --      & 9.52    & --      & --      \\
        4658037550374108288 & c      & 12.74         & 82.98 & -69.43 & 3.38    & 4.49    & --      & 27.37   & 25.88   & --      \\
                            &        &               &       & \vdots                                                             \\
        4661238159953322752 & c      & 14.86         & 77.07 & -69.03 & 6.00    & --      & --      & 41.59   & --      & --      \\
        4661238159981654912 & o      & 12.64         & 77.08 & -69.03 & 32.47   & 2.02    & --      & 14.56   & 10.32   & --      \\
        4661238159981658880 & c      & 14.24         & 77.08 & -69.04 & 407.60  & --      & --      & 45.08   & --      & --      \\
        4661238159982290816 & o      & 11.16         & 77.06 & -69.03 & 50.23   & 70.37   & 2.02    & 28.86   & 23.21   & 15.64   \\
        4661238159982290816 & c      & 11.16         & 77.06 & -69.03 & 90.99   & 2.01    & 3.83    & 46.73   & 31.03   & 22.84   \\
        \bottomrule
    \end{tabular}
    \caption{Results of the prewhitening for the ATLAS light curves. Here, camera refers to the pass-band of the photometry (orange and cyan). Note that some targets have detected periodicity in one camera but not the other. \rev{Here, the amplitudes are in parts-per-thousand (ppt).}}
    \label{tab:results}
\end{table*}

The ATLAS PL relation derived here is different from that of the OGLE relation in two notable ways. Firstly, the edge of the PL relation for the longest period variables (commonly known as the Long Secondary Period sequence; (LSP; \citealt{Hinkle2002Velocity, Olivier2003Origin, Nicholls2009Long, Stothers2010Giant, Takayama2015Testing, Takayama2020Long, Takayama2023Period}) is slightly warped. This is because ATLAS observations in the Southern Hemisphere commenced just over 2 years ago, so that the baseline of the photometry used here is insufficient to probe the longest periods. The next feature of note is the intermediate sequence E, comprising multiple star systems, lying between the LSP and Mira sequence. For the OGLE data, this sequence is only slightly resolved. Here, we observe a long continuous sequence extending to fainter stars.

\subsection{Amplitude ratios}

\begin{figure}
    \centering
    \includegraphics[width=\linewidth]{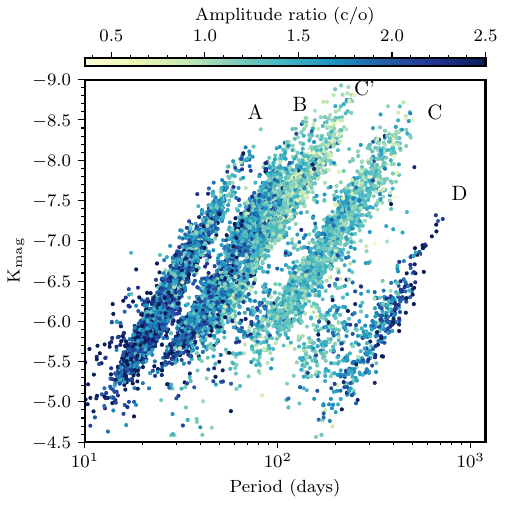}
    \caption{The LMC PL relation colored by the ratio of the dominant period amplitudes measured in the ATLAS cyan and orange passbands.}
    \label{fig:ampratio}
\end{figure}

Multicolor photometry is a significant advantage in asteroseismology. This is because the observed amplitude ratios deduced for light curves observed in different pass-bands have been used to identify the angular mode degree $l$ of main-sequence pulsators \citep[e.g.][]{Breger2000Scuti,Dupret2003Photometric,Aerts2004Longterm,2017EPJWC.15204001D,Brunsden2018Frequency}. \rev{We calculate the amplitude ratio of modes with the same periods in both ATLAS passbands (cyan / orange) for our sample in Figure~\ref{fig:ampratio} The subset of stars depicted in the PL diagram represents those measured by ATLAS and cross-matched with OGLE data, where the absolute difference between the measured ATLAS and OGLE dominant periods is less than 1 day. This criterion ensures the cleanest possible sample for comparison.}

The most striking feature of this comparison is the large amplitude ratio of the Mira sequence C (and adjacent C'). This is a well-known result from earlier studies of Mira variables, where the amplitudes of variation decrease with the wavelength (\citealt{Smith2002Infrared} and references therein).

Previous work has indicated the existence of a sequence F lying between sequences C' and C \cite{Soszynski2013Semiregular}. This sequence is typified by small amplitudes and poorly defined periods. Many stars existing on this sequence have additional secondary periods lying between sequences C and D. We see some evidence for this sequence in Fig.~\ref{fig:ampratio}, as a patch of variability with high amplitude ratios which clearly do not belong to either sequence C' or C. This is also apparent in the pre-cleaned sample of variables in Fig.~\ref{fig:pl}. Most of the stars with primary periods existing here are multi-periodic according to the ATLAS photometry.

\section{Conclusion}

In this paper we provide the first measurement of the period-luminosity relation of long-period variables in the Large Magellanic Clouds using photometry from the Asteroid Terrestrial-impact Last Alert System (ATLAS). Our main results are as follows:
%have laid the foundations for a future all-sky Milky Way map using long-period variables with ATLAS photometry. To estimate distances for these stars requires separating out individual sequences of the PL relation based on their amplitude, or incorporating the amplitude into the distance estimation itself (see, e.g., \citealt{Hey2023Far}). We have;
\begin{enumerate}
    %\item Isolated 631,259 M-giant stars in the Large Magellanic Cloud, and applied a blind search for periodic signals to the ATLAS photometry.
    \item We establish flux conversions between the ATLAS orange and cyan pass-bands to Gaia and $ugriz$ photometry for cool giant stars accurate to 0.02 and 0.03 mag for ATLAS orange and cyan bands respectively. In crowded fields, these relations can be used in combination with difference photometry to recover the flux variation of the signal.
    \item \rev{We selected 631,259 M-giant stars in the Large Magellanic Cloud, and applied a blind search for periodic signals to the ATLAS photometry. We identified long-period variability in 257,717 unique stars in the LMC. After cleaning the sample, we recover 44,459 stars which demonstrate the known PL sequences.}
          % \item 
    \item We identified that sequences C' and C in the PL relation have similar amplitude ratios in the dominant periods. This implies that these sequences are of identical angular degree.
\end{enumerate}

The ATLAS P-L relations derived here are laying the foundation for measuring an all-sky distance map of the Milky Way using long-period variables with ATLAS photometry. Estimating distances for these stars requires separating out individual sequences of the PL relation based on their amplitude, or incorporating the amplitude into the distance estimation itself (see, e.g., \citealt{Hey2023Far}). We demonstrate that we can almost completely recover the full sequence of PL relations with less than 2 years worth of photometric data, which will further improve as ATLAS observations continue.

To facilitate future analysis of this data-set, we make available all light curves processed throughout this work. This includes photometry of stars that were removed in the LMC cut (Sec.~\ref{sec:cuts}), as well as targets for which no significant period of variability was detected. These light curves are supplied with none of the quality cuts used in this paper, however, we also include a script to reproduce the light curves in their state presented herein. \rev{These light curves are available at Zenodo in csv format.}

\rev{Our work represents a comprehensive analysis of the well-known LPVs in the LMC with another instrument (ATLAS). We demonstrate that we can almost completely recover the full sequence of PL relations with less than 2 years worth of photometric data. The distribution of amplitudes in the ATLAS passband will be instrumental in exploiting the PL relation for distance estimation.}

\section*{acknowledgements}

    D.R.H. and D.H. acknowledge support from NSF (AST-2009828) and NASA (80NSSC24K0621). D.H. also acknowledges support from the Alfred P. Sloan Foundation and the Australian Research Council (FT200100871).
    This work has made use of data from the Asteroid Terrestrial-impact Last Alert System (ATLAS) project. The Asteroid Terrestrial-impact Last Alert System (ATLAS) project is primarily funded to search for near earth asteroids through NASA grants NN12AR55G, 80NSSC18K0284, and 80NSSC18K1575; byproducts of the NEO search include images and catalogs from the survey area. This work was partially funded by Kepler/K2 grant J1944/80NSSC19K0112 and HST GO-15889, and STFC grants ST/T000198/1 and ST/S006109/1. The ATLAS science products have been made possible through the contributions of the University of Hawaii Institute for Astronomy, the Queen’s University Belfast, the Space Telescope Science Institute, the South African Astronomical Observatory, and The Millennium Institute of Astrophysics (MAS), Chile.  
    The Shappee group at the University of Hawai`i is supported with funds from NSF (grants AST-1908952, AST-1911074, \& AST-1920392) and NASA (grants HST-GO-17087, 80NSSC24K0521, 80NSSC24K0490, 80NSSC24K0508, 80NSSC23K0058, \& 80NSSC23K1431).
    D.H. acknowledges support from the Alfred P. Sloan Foundation, the National Aeronautics and Space Administration (80NSSC21K0652), the National Science Foundation (AST-2009828) and the Australian Research Council (FT200100871).
    \rev{The national facility capability for SkyMapper has been funded through ARC LIEF grant LE130100104 from the Australian Research Council, awarded to the University of Sydney, the Australian National University, Swinburne University of Technology, the University of Queensland, the University of Western Australia, the University of Melbourne, Curtin University of Technology, Monash University and the Australian Astronomical Observatory. SkyMapper is owned and operated by The Australian National University's Research School of Astronomy and Astrophysics. The survey data were processed and provided by the SkyMapper Team at ANU. The SkyMapper node of the All-Sky Virtual Observatory (ASVO) is hosted at the National Computational Infrastructure (NCI). Development and support of the SkyMapper node of the ASVO has been funded in part by Astronomy Australia Limited (AAL) and the Australian Government through the Commonwealth's Education Investment Fund (EIF) and National Collaborative Research Infrastructure Strategy (NCRIS), particularly the National eResearch Collaboration Tools and Resources (NeCTAR) and the Australian National Data Service Projects (ANDS). The DOI of the SkyMapper DR2 release used in this paper is 10.25914/5ce60d31ce759.}
    \rev{"This publication makes use of data products from the Two Micron All Sky Survey, which is a joint project of the University of Massachusetts and the Infrared Processing and Analysis Center/California Institute of Technology, funded by the National Aeronautics and Space Administration and the National Science Foundation."}

\vspace{5mm}
% \facilities{HST(STIS), Swift(XRT and UVOT), AAVSO, CTIO:1.3m,
% CTIO:1.5m,CXO}

\software{astropy \citep{astropycollaborationAstropyCommunityPython2013, price-whelanAstropyProjectBuilding2018},
    numpy \citep{harris2020array},
    scipy \citep{Jones2001SciPy},
    scikit-learn \citep{scikit-learn},
    matplotlib \citep{Hunter2007Matplotlib}
}
% \appendix

\begin{singlespace}
    \bibliographystyle{aasjournal}
    \bibliography{lib}
\end{singlespace}

\end{document}

%% file: commands.tex
\usepackage{xspace,amsmath,amssymb}
\usepackage{tabularx}
\usepackage[LGR,T1]{fontenc}

\usepackage[utf8]{inputenc}
\usepackage{textcomp} % provide euro and other symbols
\usepackage[capitalize]{cleveref}
\definecolor{xlinkcolor}{cmyk}{1,1,0,0}

%% file: authors.tex
\correspondingauthor{Daniel Hey}
\email{dhey@hawaii.edu}
\author[0000-0003-3244-5357]{Daniel Hey}
\affiliation{Institute for Astronomy, University of Hawai`i, 2680 Woodlawn Drive, Honolulu, HI 96822, USA}

\author[0000-0003-2858-9657]{John Tonry}
\affiliation{Institute for Astronomy, University of Hawai`i, 2680 Woodlawn Drive, Honolulu, HI 96822, USA}

\author[0000-0003-4631-1149]{Benjamin Shappee}
\affiliation{Institute for Astronomy, University of Hawai`i, 2680 Woodlawn Drive, Honolulu, HI 96822, USA}
    
\author[0000-0001-8832-4488]{Daniel Huber}
\affiliation{Institute for Astronomy, University of Hawai`i, 2680 Woodlawn Drive, Honolulu, HI 96822, USA}
\affiliation{Sydney Institute for Astronomy (SIfA), School of Physics, University of Sydney, NSW 2006, Australia}